\documentclass[aps,prl,twocolumn,a4paper,10pt,notitlepage,superscriptaddress,showpacs]{revtex4-1}%

\usepackage{amssymb}
\usepackage{amsmath}
\usepackage{graphicx}
\usepackage[font=scriptsize,justification=raggedright]{caption}
\usepackage{bm}
\usepackage[dvipsnames]{xcolor}

\begin{document}

\author{Angelo Rosa}
\affiliation{SISSA (Scuola Internazionale Superiore di Studi Avanzati), Via Bonomea 265, 34136 Trieste, Italy}
\email{anrosa@sissa.it}
\author{Jan Smrek}
\affiliation{Faculty of Physics, University of Vienna, Boltzmanngasse 5, A-1090 Vienna, Austria}
\email{jan.smrek@univie.ac.at}
\author{Matthew S. Turner}
\affiliation{Department of Physics and Centre for Complexity Science, University of Warwick, Coventry, CV4 7AL, UK}
\affiliation{Department of Chemical Engineering, Kyoto University, Kyoto, Japan} 
\email{M.S.Turner@warwick.ac.uk}
\author{Davide Michieletto}
\affiliation{School of Physics and Astronomy, University of Edinburgh, Peter Guthrie Tait Road, Edinburgh, EH9 3FD, UK}
\affiliation{MRC Human Genetics Unit, Institute of Genetics and Molecular Medicine, University of Edinburgh, Edinburgh EH4 2XU, UK}
\affiliation{Department of Mathematical Sciences, University of Bath, North Rd, Bath BA2 7AY, UK}
\email{davide.michieletto@ed.ac.uk}

\title{Threading-Induced Dynamical Transition in Tadpole-Shaped Polymers}

\begin{abstract}
The relationship between polymer topology and bulk rheology remains a key question in soft matter physics.
Architecture-specific constraints (or threadings) are thought to control the dynamics of ring polymers in ring-linear blends,
which thus affects the viscosity to range between that of the pure rings and a value larger, but still comparable to, that of the pure linear melt. 
Here we consider qualitatively different systems of linear and ring polymers, fused together in ``chimeric'' architectures.
The simplest example of this family is a ``tadpole''-shaped polymer -- a single ring fused to the end of a single linear chain.
We show that polymers with this architecture display a threading-induced dynamical transition that substantially slows chain relaxation.
Our findings shed light on how threadings control dynamics and may inform design principles for chimeric polymers with topologically-tunable bulk rheological properties.
\end{abstract}

\maketitle

\paragraph{Introduction -- }
The tube and reptation theories underpin our understanding of complex fluids~\cite{DoiEdwardsBook,DeGennesBook1979}. However, the seemingly innocuous joining of the polymers' ends to form rings poses a problem that has been puzzling the polymer physics community for over three decades~\cite{Edwards1967,Roovers1988,Klein1986,Cates1986,Rubinstein1986,Muller2000,McLeish2002, Ferrari2002,Kapnistos2008,HalversonJCP2011_1,HalversonJCP2011_2,Sakaue2011,RosaEveraersPRL2014,Vlassopoulos2016,Ge2016,MichielettoSoftMatter2016,Sakaue2018,Schram2019,SmrekKremerRosa2019,Soh2019}.
How do topology-specific constraints affect the static and dynamic properties of a dense solution of such polymers?

Entangled solutions of pure unlinked ring polymers can now be synthesised~\cite{Kapnistos2008,Doi2015}. However, the presence of even a small fraction of linear contaminants dramatically slows their dynamics through ring-linear interpenetration~\cite{Robertson2007,Kapnistos2008,Chapman2012,HalversonPRL2012,Zhou2019}.
This slowing down shares some similarities with the one computationally discovered in systems of pure rings~\cite{MichielettoACSML2014,Lee2015,Tsalikis2016,Smrek2016}, where inter-ring threadings drive a ``topological glass'' state due to a hierarchical network of threadings -- ring-specific topological constraints~\cite{Lo2013,MichielettoPNAS2016,MichielettoPRL2017,MichielettoPolymers2017,SmrekNatComm2020}.
In ring-linear blends the linear chains cannot set up a hierarchical network of constraints and the rings are thus bound to relax on time-scales comparable to the reptative disengagement of the linear chains~\cite{Mills1987,Roovers1988,Tead1992,ParisiRubinsteinMacromolecules2020} which perform most of the threadings:
this limits severely any opportunities for further tuning of bulk rheology by using pure mixtures of ring and linear chains.

%
\begin{figure}
        \centering
        \includegraphics[width=0.45\textwidth]{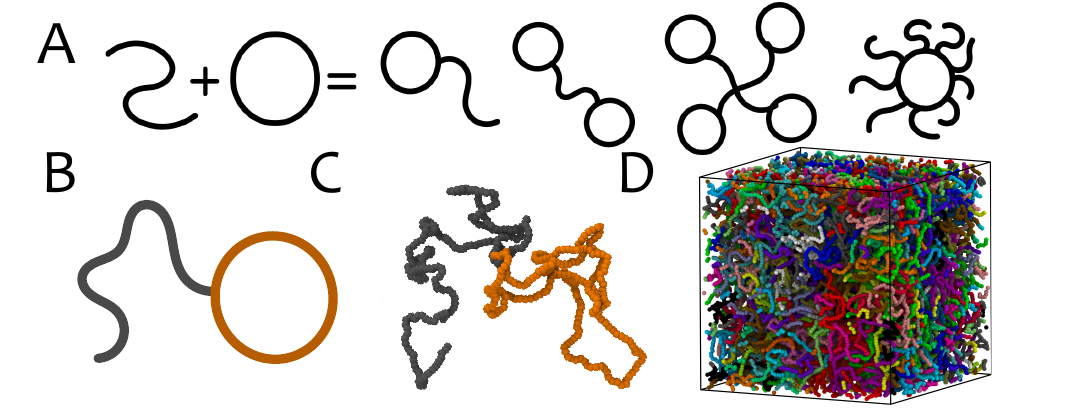}
        \caption{\textbf{A} Chimeric polymers from ring and linear chains fused together. \textbf{B} Tadpole-shaped polymers are the simplest such chimeric structure, shown as a schematic with orange ``head'' and grey ``tail''. \textbf{C} Typical simulated conformation of a tadpole and \textbf{D} an equilibrated system of $80$ tadpoles. Here the \underline{c}ircular and \underline{l}inear sections both have 250 monomers, written $(C,L)=(250,250)$.}
        \vspace{-0.8 cm}
        \label{fig:snap}
\end{figure}

To overcome this limitation, and inspired by
quickly progressing technical advances in topological polymer synthesis~\cite{Tezuka2001,Suzuki2014,Tomikawa2016},
here we investigate the behaviour of polymer architectures that simultaneously display linear and unknotted and unlinked circular topologies. We dub these architectures ``chimeric'' -- the name given to any mythical animal formed from parts of various other animals  (Fig.~\ref{fig:snap}A). The simplest example of a chimeric architecture is that of a tadpole-shaped polymer -- ``tadpole'' for brevity (see Fig.~\ref{fig:snap}B-C) -- which has recently been realised experimentally~\cite{Doi2015a,Polymeropoulos2019}
and has attracted considerable attention in the field of protein folding~\cite{Niemyska2016,Dabrowski-Tumanski2016a}.

While a broader class of polymers (dubbed ``topological'') has been studied in dilute conditions~\cite{Uehara2016,Keesman2013}, in this Letter we focus on entangled, semi-dilute concentrations and report the first Molecular Dynamics simulation (Fig.~\ref{fig:snap}D) of tadpole-shaped polymers in this regime.
Our main finding is that we observe a dynamical transition in which systems of tadpoles with long enough tails and heads display a markedly slower dynamics than a corresponding system of linear chains with equal mass.
This extremely slow dynamics is expected to arise only at asymptotically large lengths in systems of pure rings~\cite{MichielettoPNAS2016,SmrekNatComm2020},
while it cannot be achieved in standard blends of ring and linear chains~\cite{Mills1987,Tead1992,Kapnistos2008,HalversonPRL2012}
where only a $\sim2$-fold increase in viscosity has been reported~\cite{Roovers1988,ParisiRubinsteinMacromolecules2020};
while in blends there is no strategy to slow down the linear fraction beyond their natural reptative dynamics, in tadpoles this is achieved by a system-spanning (percolating) set of topological constraints.

%
\paragraph{Tadpole Microrheology -- }
We model tadpole-shaped polymers as bead-spring chains made of a ``tail'' (linear) and a ``head'' (circular) components. The monomers are connected by finitely extensible (FENE) bonds and we impose a persistence length $l_p=5\sigma$, with $\sigma$ the size of a monomer, via a Kratky-Porod potential (see SI). The junction between head and tail is freely flexible and we consider athermal solvents in which the beads interact via a purely repulsive Lennard-Jones (WCA) potential~\cite{KremerGrestJCP1990}.
The systems are made of $M$ chains with $N$ beads each at the overall monomer density $\rho = NM/V = 0.1 \sigma^{-3}$
(about 10 times the overlap concentration).
With these choices, the corresponding entanglement length for a system of linear chains is $N_e=40$ beads~\cite{Everaers2004,RosaEveraersPRL2014}; our longest tadpoles have tails $10 N_e$ long, thus putting them well into the entangled regime.
The simulations are performed
in implicit solvent
at fixed volume and temperature by weakly coupling the dynamics of the monomers with a heat bath via LAMMPS~\cite{Plimpton1995}.
The Langevin equations are evolved using a velocity-Verlet algorithm with integration step $\Delta t=0.012 \tau_{LJ}$, where $\tau_{LJ}= \sigma(m / \epsilon)^{1/2}$ is the Lennard-Jones time (see SI).

\begin{figure}
        \centering
        \includegraphics[width=0.45\textwidth]{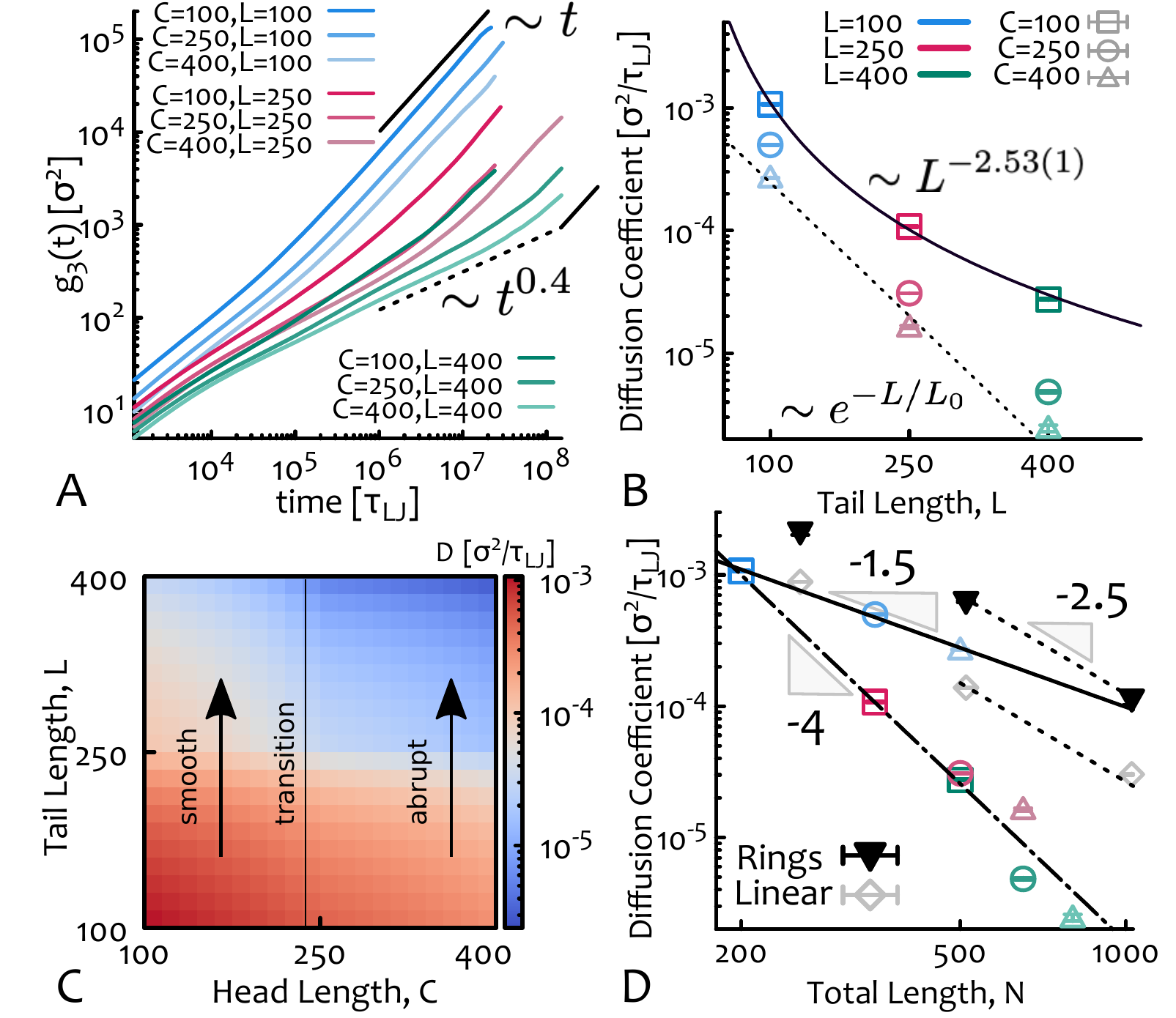}
        \caption{ \textbf{A} Mean-square displacement of the centre of mass, $g_3(t)$, of the tadpoles. \textbf{B} Log-linear plot of long-time diffusion coefficient $D$ against tail length $L$. The data set with $C=100$ is well fitted by a power law $\sim L^{-a}$ with $a=2.53(1)$ while tadpoles with larger heads display a qualitatively different slowing down with $a=a(L)$ increasing with tail size and compatible with an exponential (shown as a dashed line as a guide for the eye). \textbf{C} Interpolated heat-map of $D$ in the 2D parameter space $(C,L)$. \textbf{D} Plot of $D$ against total contour length and compared with the dynamics of pure linear and ring polymers. The solid, dashed and dashed-dotted lines are guides for the eye. The dashed line indicates the known scaling for asymptotic ring and linear chains~\cite{HalversonJCP2011_2}.
        Note that $D(L=400,C=400)$ is an upper bound value as the system has not reached free diffusion within our longest simulation runtime.
         }
        \label{fig:msd}
\end{figure}

To characterise the dynamics of the tadpoles we measure the averaged mean-square displacement (MSD) of their centre of mass (CM) as $g_{3}(t) = \left\langle ( {\vec r}_{i}(t_0+t) - {\vec r}_{i}(t_0))^2 \right\rangle$, where ${\vec r}_{i}(t)$ is the position of the CM of the $i$-th tadpole at time $t$ and $\langle\, \cdots \rangle$ indicates time and ensemble average (see Fig.~\ref{fig:msd}A). 
The trajectories display a subdiffusive regime at short-intermediate times which appears to scale as $g_3(t) \sim t^{0.4}$ for our largest tadpoles
(we compute the dynamical exponent $\alpha(t)=d\log{g_3}/d\log{t}$ in SI).
We note that this scaling exponent is distinct from, and smaller than, that of pure entangled linear chains ($t^{0.5}$) and also pure rings ($t^{0.75}$)~\cite{HalversonJCP2011_2} suggesting that tadpole dynamics appears to follow new physical mechanisms that are distinct from those of polymers with simpler topologies.

To quantify how the dynamics varies with tadpole design we compute the large-time diffusion coefficient of the centre of mass as $D=\lim_{t\rightarrow \infty} g_3(t)/6t$
({\it i.e.} we constrain the dynamical exponent $\alpha=1$ and choose a time range for which this is accurate, see SI)
and plot it as a function of tail length in Fig.~\ref{fig:msd}B. From this one should notice that the different designs display qualitatively different behaviours: for small head $C=100$ the slowing down with tail length ($L$) is well fitted by a power law $D \sim L^{-2.53(1)}$ similar to that of pure reptating linear chains~\cite{KremerGrestJCP1990,HalversonJCP2011_2} -- this suggests that the interactions between tails dominates the dynamics in this case; on the other hand, the two sets of simulations with $C = 250$ and $C = 400$ display a qualitatively different scaling behaviour whereby $D \sim L^{-a}$ with $a>3$ and increases with $L$, yielding a dynamics slower than reptation. Interestingly, comparing the square sum of residuals reveals that these two datasets are better fitted by an exponential, rather than a power law, decay. This change, or transition, in behaviour can also be qualitatively visualised in a heat-map of $D$ as a function of tadpole design $(C,L)$: $D$ decays smoothly for $C<250$ and more abruptly for $C>250$ (Fig.~\ref{fig:msd}C).

\begin{figure*}
        \includegraphics[width=1.0\textwidth]{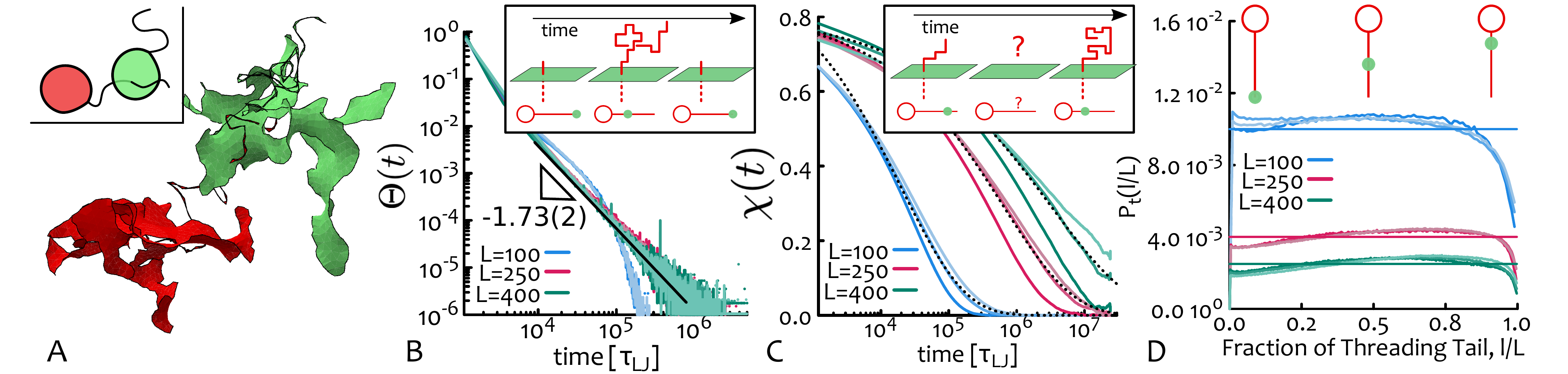}
        \caption{ \textbf{A} Snapshot of two threading tadpoles with their minimal surfaces highlighted in red and green. (Inset) Sketch of the snapshot. 
                \textbf{B} Distribution of return times $\Theta(t)$ as defined in Eq.~\eqref{eq:surv} and representative fit $\sim t^{-\beta}$ with $\beta=1.74 \pm 0.02$ for $C=250$, $L=400$. (Inset) Mapping to an anomalous Brownian walk in 1D along the tail.
        \textbf{C} Two time-point correlator $\chi(t)$. Dashed lines are representative stretched exponential fits yielding exponents $\gamma=0.359(5)$ for $C=250$,$L=400$,  $\gamma=0.416(5)$ for $C=250$,$L=250$ and $\gamma=0.459(4)$ for $C=250$,$L=100$.  (inset) Graphical sketch of the two-point correlation, stressing that $\chi(t)$ is insensitive to threading history.
                \textbf{D} Threading lengths are uniformly distributed. The horizontal lines mark inverse tail length, {\it i.e.} $1/L$, for the three sets. The distributions $P_t$ are normalised so that $\sum_{l=1}^L P_t(l)=1$.
        }
        \label{fig:threadings}
\end{figure*}

Importantly, as shown in Fig.~\ref{fig:msd}D, while the dynamics displayed by the system of tadpoles with $C=100$ interpolates in between the pure-ring and pure-linear dynamics, the two sets with $C \geq 250$ are markedly slower and they follow a qualitatively different trend also as a function of total length $N=C+L$.
Thus, our findings strongly suggest that via targeted design of tadpole structure -- and in principle other chimeric architectures --  it is possible to achieve a fine control over the bulk rheology and over a range that is orders of magnitude broader than the one that can be achieved using simpler architectures within the same window of polymer length.
It should also be highlighted that while adding linear contaminants to solutions of rings only generates systems that interpolate between the pure-ring and pure-linear behaviours~\cite{HalversonPRL2012,Kapnistos2008}, with chimeric polymers, due their fused architecture, we can produce emergent collective behaviours which have no counterpart in ring-linear blends.
We now show that these observed collective phenomena are due to inter-tadpole ``threadings'', {\it i.e.} piercing of a tadpole's tail through the head of another.

\paragraph{Threading Statistics -- }
Motivated by previous work~\cite{MichielettoACSML2014,HalversonJPA2013,Doi2015a}, we hypothesise that threadings may give rise to an emergent slowing down in our entangled tadpoles. To identify threadings we use the concept of minimal surfaces~\cite{Lang2013,Smrek2016,SmrekKremerRosa2019}: we first fix a boundary using the position of the beads forming the heads and generate an initial triangulated surface; we then evolve this surface via the Surface Evolver under the action of surface tension until the area is minimised~\cite{Brakke1992}. Once a minimal surface is defined per each tadpole head, we look for intersections between all possible pairs of tail and head surface (see Fig.~\ref{fig:threadings}A). [We choose to exclude self-intersections as they may be ill-defined in some cases]. This strategy allows us to define a time-dependent threading matrix as follows: $T_{ij}(t) = 1$ if tadpole $j$ is threading tadpole $i$ ($i \neq j$) and $0$ otherwise.

Threadings are stochastic events that last for a certain time and we quantify the distribution of these threading lifetimes via the following quantity
\begin{equation}
\Theta(t) = \langle P(T_{ij}(t)=0 | T_{ij}(0)=1, \dots, T_{ij}(t-1)=1) \rangle
\label{eq:surv}
\end{equation}
where $P(X|Y_1, \dots, Y_n)$ is the probability of observing $X$ conditioned on $Y_1, \dots, Y_n$ being observed and $\langle \rangle$ indicates the ensemble and time average. In practice, Eq.~\eqref{eq:surv} counts the threadings with life-time exactly $t$ and the resulting curves are reported in Fig.~\ref{fig:threadings}B. To discuss these curves, we should note that Eq.~\eqref{eq:surv} calculation can be mapped to that of a first return time (or first passage time) of a Brownian Walk in 1D. In this framework, the walker represents the intersection of the tail through the head-spanning minimal surface; the walker moves along the tail as the threading diffuses in and out the minimal surface (see inset of Fig.~\ref{fig:threadings}B). The distribution of return times of a Brownian Walk is expected to be a power law and to scale as $\sim t^{\alpha/2-2}$ where $\alpha$ is the anomalous exponent of the walk~\cite{Molchan1999,Metzler2014a}. In our case the tails are expected to follow a Rouse dynamics -- confirmed by direct tracking of the piercing segment, which yields $\alpha = [0.4, 0.6]$ (see SI) -- and we thus predict the distribution of return times to scale with an exponent $\alpha/2-2 = [1.7, 1.8]$ in very good agreement with our best fits of $\Theta(t)$ for $L \geq 250$ (see Fig.~\ref{fig:threadings}B). [The curves with $L=100$ display a scaling exponent closer to $-1.5$ as their Rouse regime is shorter than our sampling time].

Importantly, we note that the slowest return time displayed by $\Theta(t)$ is still $\sim$10-fold faster than the longest relaxation of the tadpoles ($10^{6} \tau_{LJ}$ versus $10^{7} \tau_{LJ}$, compare the curves $\Theta$ with the crossover time to diffusion of $g_{3}$). This suggests that it is collective multi-threading events that control the long-time dynamics of tadpoles.

\begin{figure}
        \includegraphics[width=0.45\textwidth]{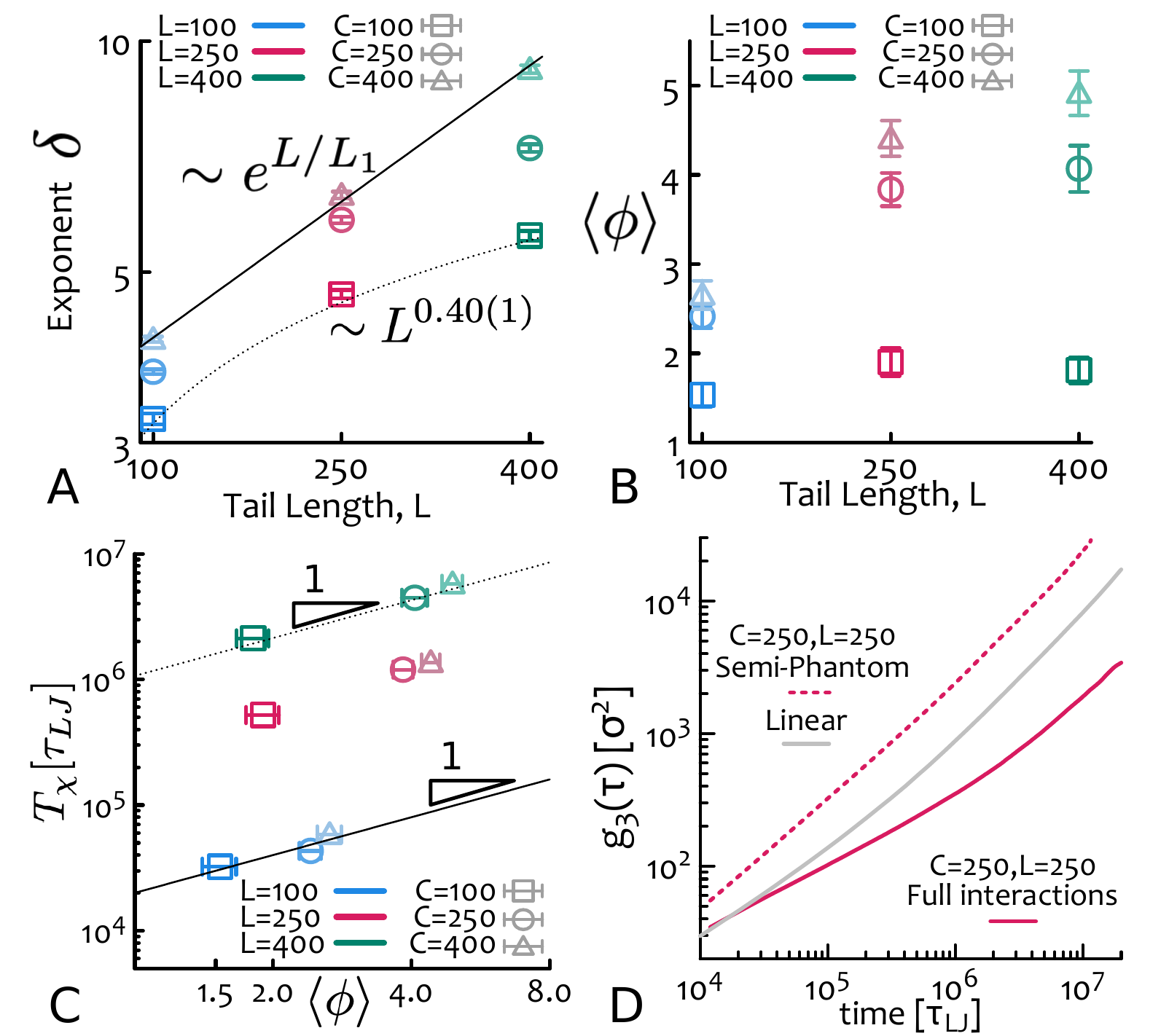}
        \caption{ 
        \textbf{A} The threading relaxation exponent $\delta$ increases with tail length as a power law $\delta \sim L^{0.40(1)}$ for small heads and exponentially $\delta \sim e^{L/L_1}$ with $L_1=367(13)$ for large heads.
                \textbf{B} Average number of threading tails per tadpole $\langle \phi \rangle$ as a function of tail length. 
                 \textbf{C} Threading correlation time $T_\chi$ scales linearly with $\langle \phi \rangle$ (with a prefactor proprtional to $L^3$) suggesting that a serial release of $\langle \phi \rangle$ threadings is needed before all constraints are released.
        \textbf{D} Comparison of $g_3(t)$ in presence and absence of threading constraints (see text).
        }
        \label{fig:thread_dynamics}
\end{figure}

In light of this we study the two time-points correlator $\chi(t) =  \langle T_{ij}(t) T_{ij}(t+t_0)) \rangle - p_T$, where $p_T=\langle \phi \rangle/(M-1)$ is the background probability that any two tadpoles are threading at any given time and $\langle \rangle$ is the average over times $t_0$ and pairs of tadpoles $(i,j)$. We note that the longest relaxation time of $\chi(t)$, {\it i.e.} the time at which $\chi \simeq 0$, broadly agrees with the crossover time to free diffusion of the tadpoles (compare Fig.~\ref{fig:thread_dynamics}C with Fig.~\ref{fig:msd}A).
This quantity is akin to a stress relaxation in polymeric systems and informs us on the relaxation dynamics of inter-tadpole threadings. By assuming that threadings are monodisperse in length we would expect $\chi(t) \sim e^{-t/T(l)}$ where $T(l)$ is the typical relaxation time of a threading of length $l$. Instead we find that $\chi(t)$ decreases as a stretched exponential $\chi(t) \sim \exp{(-A t^{\gamma})}$ as expected for a polydisperse solution of entangled linear polymers~\cite{DeGennes2002}. In the case of polymer lengths that follow a Poisson distribution, the exponent $\gamma$ can be computed via a saddle point approximation to be $\gamma = 1/(1+\beta)$ (where $\beta=2$ or $3$ for Rouse and reptation respectively)~\cite{DeGennes2002,Cates1987}. In our case, we find that the distribution of threading lengths, {\it i.e.} the portion of tail from the piercing point to the end of the tail, is instead uniform, {\it i.e.} $P(l)\sim 1/L$ (see Fig.~\ref{fig:threadings}D). Thus, to compute their relaxation we must calculate 
$\chi(t)= (1/L)\int_0^L e^{-t/T(l)} dl $,
where $T(l)= \tau_0 l^\delta$ now depends on the threading length $l$ through a generic exponent $\delta$. This function 
can be computed numerically as a function of $\tau_0$ and $\delta$ for different choices of $C$ and $L$. As expected, we find that $\tau_0$ is overall independent of either $C$ or $L$ (see SI); on the other hand, we find that $\delta$ -- which is also expected to be insensitive of $L$ within the classic reptation dynamics -- increases as a power law of $L$ for small heads and exponentially in $L$ for large heads (Fig.~\ref{fig:thread_dynamics}A). This implies that $T(l)$ diverges even more strongly than an exponential in the asymptotic limit of large tadpoles. We should note that the distinct behaviour of $T(l)$ for small and large heads mirrors the qualitatively distinct regimes observed in the decay of $D$ (Fig.~\ref{fig:msd}B). This strongly suggests that threadings play a key role in the dynamics.

The results shown up to now suggest that tadpoles with large heads have a qualitatively different dynamics with respect to the ones with smaller head; in particular, they display a much stronger slowing down and slower threading relaxation.
To explain this finding we note that the head-spanning minimal surfaces scale linearly~\cite{Smrek2016} with head length $C$ (see also SI) and, hence, tadpoles with larger heads are expected to accommodate more threadings.
In particular, we expect that the number of threadings per head should scale linearly with $C$ (and hence with $N$) in the asymptotic limit.  From the time-dependent threading matrix $T_{ij}(t)$ we can extract the mean number of (passive) threadings per tadpole as $\langle \phi \rangle \equiv \langle \sum_{j \neq i} T_{ij}(t) \rangle$, where the average is performed over time and tadpoles.
This quantity is reported in Fig.~\ref{fig:thread_dynamics}B and indeed it shows that for small heads the number of threadings is saturated at modest tail lengths; on the other hand, larger heads can accommodate up to 5 threadings, on average, and often each threading is made by more than one piercing (see SI). Importantly, they appear to saturate at much larger values of tail length and arguably will scale extensively with $L$ in the limit of large heads $C$.
A natural consequence of the fact that $\langle \phi \rangle > 1$ is that these systems are percolating, {\it i.e.} the largest number of tadpoles connected by threadings is comparable with the system size. In particular we find that the critical threading length required to set up a percolating cluster of tadpoles is $l_c/L =1/ \langle \phi \rangle$ (see details in SI).

To correlate the mean number of threadings with a dynamical quantity we extract a characteristic time from $\chi$ as $T_\chi=\int_0^{\infty} \chi(t) \, dt$ and find that $T_\chi \sim \langle \phi \rangle$ (Fig.~\ref{fig:thread_dynamics}C) suggesting that the full relaxation of threading constraints depends on the number of threadings. This can be explained by noting that the full relaxation appears to need $\langle \phi \rangle$ serial release events before (all) the threading constraints are released. We also note that the diffusion coefficient strongly depends on the mean threading number (see SI). An exact quantification of the variation of tadpole mobility with number of threadings alone is difficult since $D$ is also a function of total contour length. 

To unambiguously detect the role played by threadings in the dynamics of tadpoles we thus propose a new strategy: we investigate a symmetric ({\it i.e.} $C=250$, $L=250$) system of tadpoles with phantom (no steric) interactions between heads and tails, while maintaining standard self-avoidance between pairs of monomers belonging to two heads or two tails. This entails that threadings of heads by tails are no longer topological constraints for the dynamics of the tadpoles. In order to fairly compare with our other results we compress this system $2-$fold (in volume) in order to maintain the effective (self-avoiding) monomer density at $\rho=0.1 \sigma^3$. We find that the absence of effective threading results in a much faster transition to free diffusion and a 14-fold enhancement of diffusion coefficient (Fig.~\ref{fig:thread_dynamics}D). This finding provides independent and unambiguous evidence that it is indeed the threadings between chains that are responsible for their correlated (subdiffusive) dynamics over short-intermediate times and resulting retarded centre-of-mass diffusion. We note that in dilute conditions, the dynamics of tadpoles does not depend on their design; this further confirms that the observed behaviour is due to collective interactions (see SI Fig.~S11).

Finally, we mention that our results are in fair quantitative agreement with experiments~\cite{Doi2015a} (see SI) and that the zero-shear viscosity obtained from both, experimental and simulated tadpoles, are best fitted by a power law with exponent close to $\eta_0 \sim L^{4.5}$. Nonetheless, the data also suggest that both experiments and simulations are performed in a crossover regime and our analysis strongly supports the argument that in the asymptotic regime the tadpoles' mobility should slow down exponentially in tail length (Fig.~\ref{fig:msd}B and \ref{fig:thread_dynamics}).

\paragraph{Conclusions -- }
In this work we have investigated the dynamics of entangled systems of tadpole-shaped polymers, as the simplest example of a broader family of ``chimeric'' polymers formed by the combination of unknotted and unlinked loops and branches (Fig.~\ref{fig:snap}A). While similar architectures had been investigated in the dilute regime~\cite{Uehara2016,Keesman2013}, here we design entangled systems with the aim of understanding how to achieve a fine control over threading topological constraints and, in turn, over the rheology of the bulk.

Here we have discovered that it is possible to design polymer architectures that can span a much larger dynamical range than that achievable with simpler architectures at fixed polymer mass. For instance, using tadpole-shaped polymers, we can explore a dynamical range that is about two orders of magnitude broader than the one for linear chains with modest lengths $N/N_e=25$ (Fig.~\ref{fig:msd}D).
Importantly, this phenomenon cannot be reproduced using ring-linear blends as their slowing down due to threading was observed to be of order unity
compared with that of linear chains only~\cite{Tead1992,Roovers1988,HalversonPRL2012,Kapnistos2008}
and expected to scale only linearly with rings mass~\cite{ParisiRubinsteinMacromolecules2020}. 

We argue that this marked difference is due to a lack of a strategy to slow down linear chains more than reptation in ring-linear blends. On the contrary, the fused architecture of tadpoles (and of higher order exotic polymers) together with the emergence of a hierarchical, percolating set of threading topological constraints, entails that the process of constraint release imposed by linear tails on circular heads propagates back to tails too, causing a dramatic and system-wide slowing down. We feel it would be very interesting to directly compare the dynamics of tadpoles and that of ring-linear blends with same values of $C$ and $L$ in simulations and experiments.

By using minimal surfaces (Fig.~\ref{fig:threadings}) and semi-phantom interactions (Fig.~\ref{fig:thread_dynamics}D) we unambiguously demonstrated that inter-tadpole threadings play a major role in the dynamics and that this effect is
not due to single threadings (Fig.~\ref{fig:threadings}B)
but to correlated (Fig.~\ref{fig:threadings}C) and collective (Fig.~\ref{fig:thread_dynamics}C) ones. Interestingly, the more the threadings per tadpole, the slower is their full relaxation (Fig.~\ref{fig:thread_dynamics}C), thus entailing further non-linear slowing down in the large $N$ limit (Fig.~\ref{fig:thread_dynamics}B).   

We have also showed that the relaxation of threadings can be mapped to that of a polydisperse system of polymers, with the caveat that here the distribution of threading lengths is uniform (Fig.~\ref{fig:threadings}D) and that the exponent of the longest relaxation time increases with $L$ (Fig.~\ref{fig:thread_dynamics}A).
This finding is in stark contrast with simpler architectures, e.g. linear, for which the relaxation exponent is insensitive on polymer length, e.g. $\delta =3$ for reptation of polymers with any $L$.

We argue that the phenomenology observed here might be generically expected across the broader family of chimeric polymers and that further fine tuning can likely be achieved by varying the number of looped structures, as well as their relative lengths. Ultimately, we envisage using these exotic architectures to tune the dynamics of specific polymers that are expensive to synthesise in large scales. Our results suggest that even a modest polymer mass can display a broad dynamical range and this property can be harnessed to keep the costs low while achieving the desired rheology through informed polymer design.
Our work might therefore serve to motivate future theoretical and experimental characterisations of entangled solutions of higher-order chimeric structures which may be
now feasibly
realised via synthetic chemistry~\cite{Tezuka2001,Doi2015,Polymeropoulos2019}
or DNA origami.

%
\paragraph{Acknowledgements.}
The authors would like to acknowledge the contribution, networking support and STSM Grant by the ``European Topology Interdisciplinary Action'' (EUTOPIA) CA17139.
This project has also received funding from the European Union's Horizon 2020 research and innovation programme under grant agreement No. 731019 (EUSMI).
DM acknowledges the computing time provided on the supercomputer JURECA at J\"ulich Supercomputing Centre.
JS acknowledges the support from the Austrian Science Fund (FWF) through the Lise-Meitner Fellowship No. M 2470-N28.
JS is grateful for the computational time at Vienna Scientific Cluster.

\bibliography{biblio}

\end{document}